\documentclass[journal,twoside,print]{ieeecolor}
\usepackage{cite}
\usepackage{float}
\usepackage{upgreek}
\usepackage{generic}
\usepackage{cite}
\usepackage{amsmath,amssymb,amsfonts}
\usepackage{algorithmic}
\usepackage{graphicx}
\usepackage{textcomp}

\usepackage{marvosym}
\def\BibTeX{{\rm B\kern-.05em{\sc i\kern-.025em b}\kern-.08em
    T\kern-.1667em\lower.7ex\hbox{E}\kern-.125emX}}
\begin{document}
\title{A Polarization and Radiomics Feature Fusion Network for the Classification of Hepatocellular Carcinoma and Intrahepatic Cholangiocarcinoma}
\author{Jia Dong\textsuperscript{*}, Yao Yao\textsuperscript{*}, Liyan Lin, Yang Dong, Jiachen Wan, Ran Peng, Chao Li\textsuperscript{\Letter} and Hui Ma\textsuperscript{\Letter} 
\thanks{Both authors contributed equally: Jia Dong; Yao Yao.}
\thanks{Corresponding authors\textsuperscript{\Letter}: Chao Li (lichao3501@163.com); Hui Ma (mahui@tsinghua.edu.cn).}
\thanks{Jia Dong is currently with Department of Statistical Science at University College London (jia.dong.23@ucl.ac.uk).}
\thanks{Jia Dong, Yang Dong, Jiachen Wan, and Hui Ma are with Guangdong Engineering Center of Polarization Imaging and Sensing Technology, Shenzhen Key Lab for Minimal Invasive Medical Technologies, Tsinghua Shenzhen International Graduate School, Tsinghua University, Shenzhen, 518055, China.
}
\thanks{Yao Yao, Yang Dong, Jiachen Wan, and Hui Ma are with the Tsinghua-Berkeley Shenzhen Institute, Tsinghua Shenzhen International Graduate School, Tsinghua University, Shenzhen, 518055, China.}
\thanks{Liyan Lin, Ran Peng, and Chao Li are with the Department of Pathology, Fujian Medical University Cancer Hospital, Fujian Cancer Hospital, Fuzhou, 350014, China.}
\thanks{This work was supported in part by National Natural Science Foundation of China (NSFC) (Grant Nos. 61527826 and 11974206) and Shenzhen Bureau of Science and Innovation (Grant No. JCYJ20170412170814624).}
}

\maketitle

\begin{abstract}
Classifying hepatocellular carcinoma (HCC) and intrahepatic cholangiocarcinoma (ICC) is a critical step in treatment selection and prognosis evaluation for patients with liver diseases. Traditional histopathological diagnosis poses challenges in this context. In this study, we introduce a novel polarization and radiomics feature fusion network, which combines polarization features obtained from Mueller matrix images of liver pathological samples with radiomics features derived from corresponding pathological images to classify HCC and ICC. Our fusion network integrates a two-tier fusion approach, comprising early feature-level fusion and late classification-level fusion. By harnessing the strengths of polarization imaging techniques and image feature-based machine learning, our proposed fusion network significantly enhances classification accuracy. Notably, even at reduced imaging resolutions, the fusion network maintains robust performance due to the additional information provided by polarization features, which may not align with human visual perception. Our experimental results underscore the potential of this fusion network as a powerful tool for computer-aided diagnosis of HCC and ICC, showcasing the benefits and prospects of integrating polarization imaging techniques into the current image-intensive digital pathological diagnosis. We aim to contribute this innovative approach to top-tier journals, offering fresh insights and valuable tools in the fields of medical imaging and cancer diagnosis. By introducing polarization imaging into liver cancer classification, we demonstrate its interdisciplinary potential in addressing challenges in medical image analysis, promising advancements in medical imaging and cancer diagnosis.
\end{abstract}


\section{Introduction}
\label{sec:introduction}
\IEEEPARstart{L}{iver} cancer is one of the most common malignant tumors worldwide. There are approximately 906000 new cases of liver cancer and 830000 liver cancer-related deaths globally each year. Hepatocellular carcinoma (HCC) accounts for 75-85\% of primary liver cancer and intrahepatic cholangiocarcinoma (ICC) for 10-15\%. Trends lead that liver cancer becomes the sixth leading cause of morbidity and the third leading cause of mortality in 2020 \cite{1}. Although HCC and ICC have different epidemiologic, etiological, and clinical characteristics, it is difficult to distinguish them completely. ICC is more invasive than HCC, indicating different treatment plans between them \cite{2}, \cite{3}. Therefore, accurate differentiation between HCC and ICC in pathological diagnosis enables clinicians to select appropriate treatment for patients, assist clinical decision-making, and improve patient prognosis and survival rate. The diagnosis of HCC and ICC is conducted by the observation and evaluation of the Hematoxylin and Eosin (H\&E)-stained combined with immunohistochemistry (IHC)-stained sections of liver pathological tissues by experienced pathologists using high resolution optical microscope clinically \cite{4,5}. In IHC detection, specific immunostaining Hep Par-1 and Arg-1 are markers for HCC, and CK19 could help to distinguish ICC from other cancers \cite{6,7}. However, for poorly differentiated cancer, the detection ability of immunohistochemistry is limited, since some poorly differentiated HCC may lose or only focally express hepatocyte specificity markers. And studies about HCCs demonstrated substantial CK19 immunostaining could lead a higher recurrence rate of cancer and higher rate of lymph node metastasis \cite{8}. Therefore, it is necessary to explore objective and rapid diagnostic methods that can effectively identify HCC and ICC without IHC.

The development of imaging techniques and artificial intelligence methods provides new ways in the classification of HCC and ICC. There are relatively few studies in distinguishing the two types of liver cancers with pathological images, whereas some related researches were carried on with computed tomography (CT) \cite{9,10,11}, ultrasound (US) \cite{12} and magnetic resonance imaging (MRI) \cite{13} images. In clinic, pathology is the golden standard of cancer diagnosis. Digital pathology has made great strides in recent years, including the digitization of pathological sections \cite{14}  and the feature extraction and analysis methods based on the digitized images \cite{15}. The former can generate high-resolution pathological images by using whole slides scanning equipment, and the latter takes Artificial Intelligence (AI) as the main image analysis method to automate the pathological diagnosis process, reduce work intensity of pathologists, and improve the objectivity and accuracy of pathological diagnosis \cite{16}. Currently, AI-based image analysis technologies for computer-aided diagnosis are becoming the core of digital pathology \cite{17}. However, challenges remain to promote wide applications of AI approaches in digital pathology. For example, it requires large amounts of high-resolution color images of stained pathological tissue sections for extracting effective pathological information, whereas the number of pathological sections from patients in hospital is limited. Therefore, to obtain more dimensional information from pathological sections and expand the variety of input data, Machine Learning (ML) models combined with optical imaging techniques were proposed for enhancing the ability for extracting structural information and assisting pathological diagnosis \cite{9,18,19}. For example, Dong et al. \cite{20} proposed a dual-modality machine learning framework for cervical intraepithelial neoplasia grading task, identifying the macro-structure and segment the target region in pathological images by deep learning-based image analysis method and then extracting the micro-structure information of the target region by emerging polarization imaging technique. 

Polarization microscopy imaging obtains specific features and effective information of pathological sections through the change of polarization state in the process of polarized light scattering propagation. Especially, it is more sensitive to the scattering by sub-wavelength scale microstructure \cite{21,22,23,24}. Mueller matrix, as a comprehensive description of the polarization-related properties of the sample, contains abundant information on microstructural and optical characterizations \cite{25}. To interpret the information encoded in the Mueller matrix, sets of polarization parameters with physical meanings were derived from the Mueller matrix by several methods such as Mueller matrix polar decomposition (MMPD), Mueller matrix transform (MMT) and so on \cite{26,27,28,29}. These polarization parameters have shown good potential in assisting diagnosis of various pathological tissues, such as liver fibrosis \cite{30,31}, breast cancer \cite{32}, skin cancer \cite{33}, and colon cancer \cite{34}. However, the existing polarization parameters have limited ability for recognition of more specific and finer pathological microstructures. Based on machine learning methods, Dong et al. \cite{35} proposed a linear discriminant analysis (LDA) based method to derive polarimetry feature parameters composed of the existing polarization parameters for quantitative characterization of specific microstructures in various breast pathological tissues. Furthermore, Liu et al. \cite{36} investigated the study about correlation degree between the polarization parameters of the breast pathological tissue samples and texture features of the corresponding pathological images. However, these two kinds of information are not only relevant, but also complementary. Specifically: It has been demonstrated that the contrast mechanism of polarization images depends on the polarization characteristics of the sample and less on the imaging resolution, enabling polarization imaging to see what is invisible in human eyes. Therefore, in the emerging polarization imaging, each pixel can be used as an independent sample with multi-dimensional polarization features, which contains high-resolution pathological features encoded in the low-resolution imaging and reveals abundant sub-wavelength scale micro-structure information at the pixel level; On the other hand, in traditional microscopic imaging, high-resolution color pathological images have rich visual information which is consistent with pathologists’ observation. The image feature parameters derived from pathological images quantify the relationship between pixels and spatial distribution and may provide macro-structural information within the target lesion area for histopathology at the image level. Image feature parameters can be automatically extracted by various data-characterization algorithms, one of which is radiomics \cite{37,38}. Radiomics is an emerging field. It can be used to fully excavate the hidden information in medical images and output a large amount image feature parameters for clinical applications such as tumor parting, treatment option, efficacy testing, and prognostic evaluation \cite{39,40}. It is also feasible to apply radiomics to the pathological images, which converts pathological image into a high dimensional feature space, including texture features, statistical features, and shape features. Overall, polarization features at the low-resolution pixel level and image features of the corresponding color pathological images at the high-resolution image level are complementary. Combining features of the two different modalities will increase the dimension of information and make the description of the tissue characteristics more comprehensive.

In this paper, we present a dual-modality ML framework designed to classify HCC and ICC within H\&E-stained pathological sections of liver tissues. This framework combines polarization features derived from Mueller matrix images with radiomics features extracted from corresponding H\&E pathological images. Differing from prior work where the discriminative feature was solely the polarization feature, the proposed fusion network encompasses two levels of fusion, conducting early fusion at the feature level and late fusion at the classification level, using both polarization features and image features as discriminative characteristics. Experimental results confirm that our proposed polarization and radiomics feature fusion network (PRFFN) outperforms single-modality machine learning classifiers in the classification of HCC and ICC. Moreover, the accuracy of the PRFFN exhibits superior robustness as imaging resolution decreases compared to radiomics feature-based classifiers. This study underscores the synergy between the polarization properties and image features of pathological samples and highlights the benefits of incorporating polarization imaging technology into contemporary image-rich digital pathology. The approach promises to offer multidimensional information and a comprehensive description of pathological samples, ultimately enhancing the accuracy and objectivity of computer-aided pathological diagnosis.

In summary, this work represents a significant advancement in the field by introducing dual-modality classification, fusing polarization features with image features to distinguish HCC and ICC in liver tissues, thus contributing to a more accurate and effective diagnostic tool for liver diseases.

\section{Methods}
\subsection{Liver Cancer Pathological Samples}
The 5-$\upmu$m-thick H\&E-stained pathological slides of liver tissues used in this study were obtained from the Fujian Medical University Cancer Hospital. The pathological samples consisted of a total of 28 slides from 28 patients, including 14 cases of HCC and 14 cases of ICC, which were removed without receiving any preoperative treatments and confirmed by pathological examination after the operation. The specimens were treated by conventional production, and each case selected a wax block taken from tumor tissue without necrosis. According to the pathological extraction specification, the size of block is about (1.5--2) cm $\times $ 1 cm $\times $ 0.2 cm, and each block was cut into slices about 5-$\upmu$m thick for regular H\&E staining. Several pathologists discussed and selected the regions-of-interest (ROIs) containing a large number of significant HCC cell structures, ICC cell structures, and non-cancerous structures in each sample, then used MATLAB Graphical User Interfaces (GUIs) to manually labelled the three structures in selected ROIs. The GUI calls the \emph{imfreehand} function to enable the expert pathologist to label the target microstructures on the H\&E image of the ROI and generates a binary image as the ground truth for the training and testing of the model. We classified the three structures by collecting and analyzing data from 148 ROIs (53 for HCC cell, 53 for ICC cell, 42 for non-cancerous) selected from 28 pathological samples. This work was approved by the Ethics Committee of Fujian Medical University Cancer Hospital.

\subsection{Data Acquisition}
\subsubsection{Experimental Setup}
Using the Mueller matrix microscope on H\&E-stained sections, we obtained the sample’s Mueller matrix image and H\&E pathological image \cite{41}.

The dual division of focal plane (DoFP) polarimeters-based Mueller matrix microscope (DoFPs MMM) is established by adding two compact modules, i.e., polarization state generator and analyzer (PSA and PSG), into a commercial transmission microscope (L2050, Guangzhou LISS Optical Instrument Co., Ltd., China), as shown in Fig. 1(a). Collimated light from the LED (633 nm, 1$\uplambda$ = 20 nm) is modulated by PSG which consists of a fixed-angle linear polarizer (P1) and a rotatable quarter-wave plate (R1) and then transmit the tissue sample. Passing through the objective lens, the scattered light is detected by PSA which consists of two 16-bit DoFP polarimeters (PHX050S-PC, Lucid Vision Labs Inc., Canada, DoFP-CCD1 and DoFP-CCD2), a 50:50 non-polarized beam splitter prism, and a fixed-angle phase retarder (R2). 

During a measurement, R1 rotates to four preset angles in order that the PSG generates four independent polarization states $S_{in}$. When R1 arrives at an angle, two DoFP-CCD conduct data acquisitions at the same time, so that a total of 4 acquisitions are carried out. The instrument matrix $A_{PSA}$ of PSA is pre-calibrated by measuring standard samples’ Mueller matrix, such as air and retarder, and calculated pixel by pixel before being applied to pathological samples. After calibration, the maximum error of the DoFPs MMM is about 1\%.
The polarization state $S_{out}$ of the outgoing light can be obtained according to:
\begin{equation}S_{out} =A_{PSA}^{-1}I .\label{eq}\end{equation}
where \emph{I} represent the intensity of the polarization component images recorded by the DoFP-CCD1 and DoFP-CCD2. Therefore, the Mueller matrix of the sample can be reconstructed by the equation:
\begin{equation}M_{sample}=\left [ S_{out} \right ] \left [ S_{in} \right ]^{-1} .\label{eq}\end{equation}
Fig. 1(b) presents the Mueller matrix measurement of the liver cancer pathological samples under a 4 $\times $ objective lens. The Mueller matrix elements are normalized by element m11 which is the intensity image. In addition, we can obtain the corresponding H\&E pathological image of sample from a color CCD under a 20 $\times $ objective lens as shown in Fig. 1(c).
\begin{figure}[H]
\centerline{\includegraphics[width=\columnwidth]{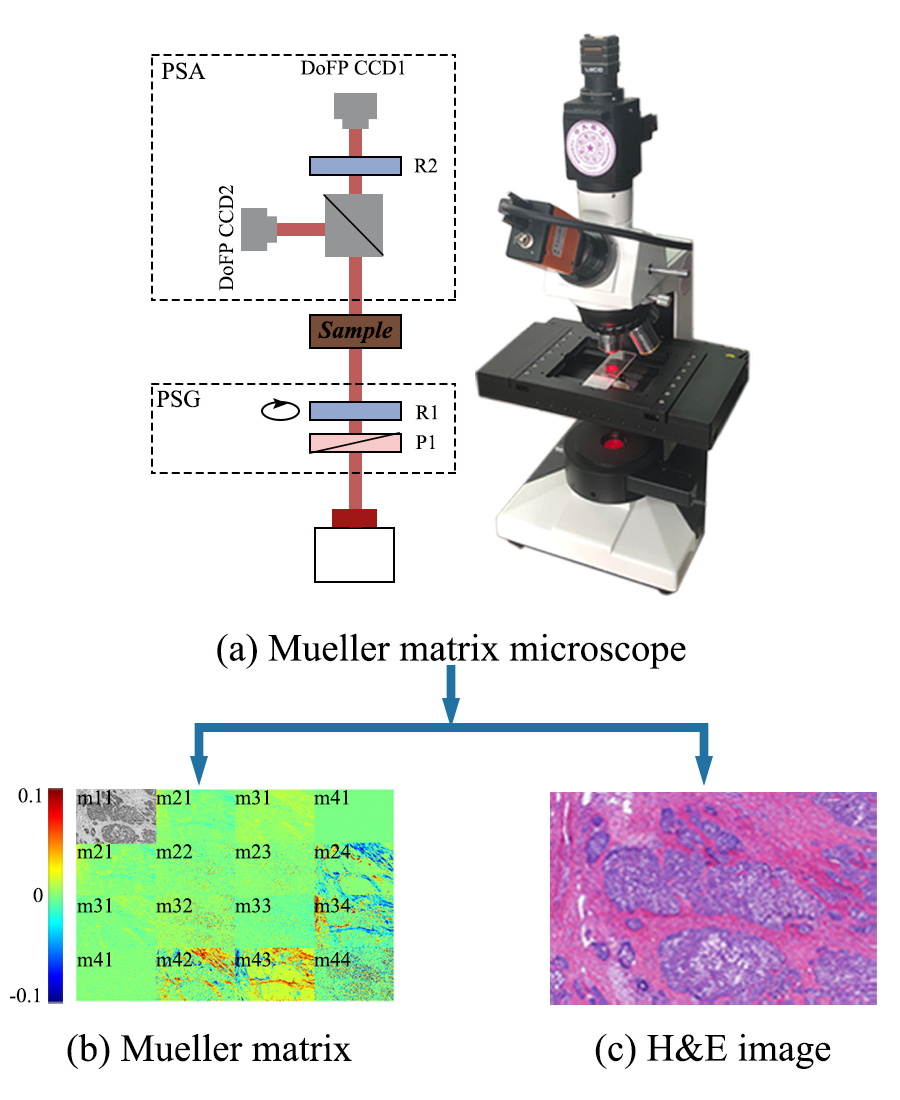}}
\caption{Data acquisition method and experimental setup. (a) Photograph and schematic of the Mueller matrix microscope. (b) An example of Mueller matrix of the pathological section of liver tissue sample under a 4$\times $ objective lens. We subtract the identity matrix from the Mueller matrix for display using the color bar ranging from -0.1 to 0.1.(c) The corresponding H\&E pathological image of sample from color CCD under a 20$\times $ objective lens.}
\label{fig1}
\end{figure}
\subsubsection{Polarimetry Basis Parameters}
\begin{figure*}[h]
\centerline{\includegraphics[width=\linewidth]{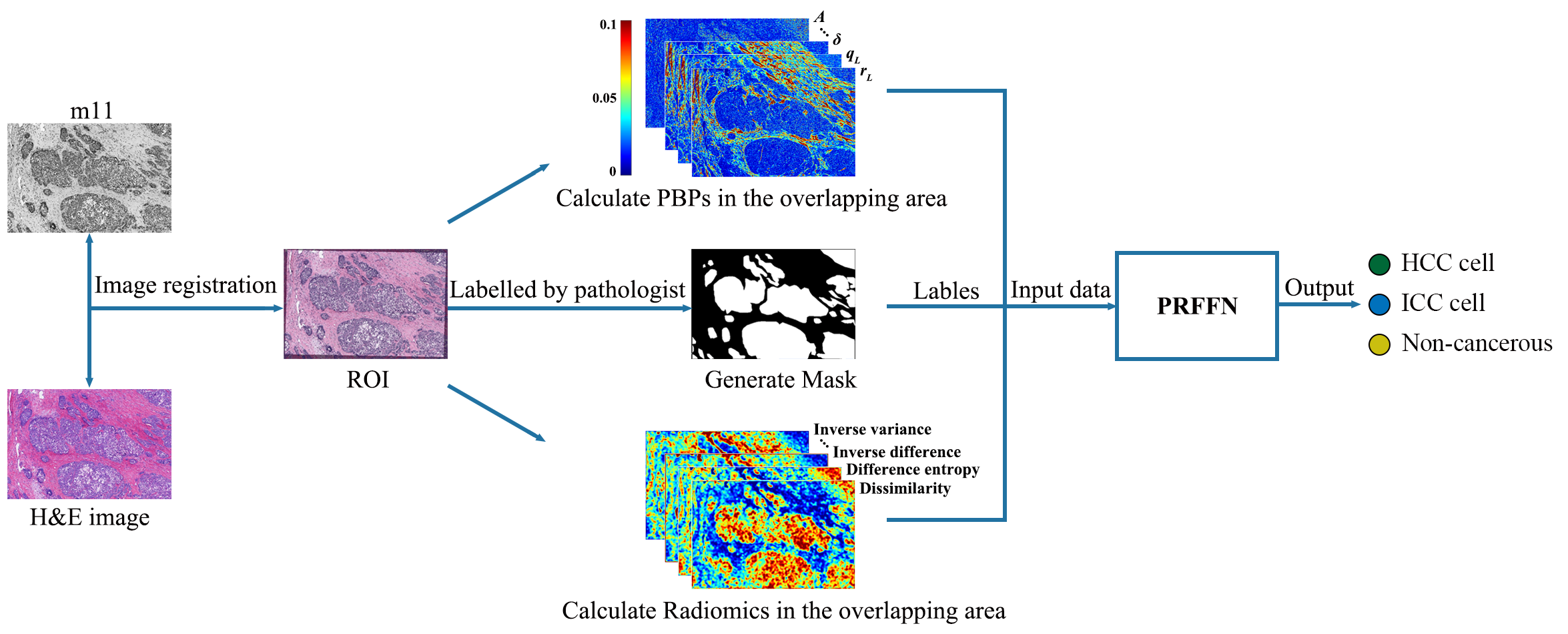}}
\caption{The framework for classifying HCC and ICC. The schematic illustrates a dual-modality method by using a fusion network which fuses polarization features derived from Mueller matrix images and radiomics features derived from the corresponding H\&E pathological images, and outlines the steps from the input of Mueller matrix image and H\&E image of the liver tissues to the output of the classification of HCC cell structures, ICC cell structures, and non-cancerous structures.}
\label{fig2}
\end{figure*}
Mueller matrix encodes complete polarization properties of
samples. Due to a lack of explicit connections between 
individual Mueller matrix elements and the microstructural
characteristics of samples, polarimetry basis parameter 
(PBPs) which have explicit physical meanings were decoded 
from the Mueller matrix to characterize the 
microstructure. MMPD method proposed by Lu and Chipman 
\cite{26} derived linear retardation $\updelta $, 
depolarization $\Delta $, optical rotation $\upvarphi $, 
and diattenuation \emph{D}. In our previous studies, we 
proposed MMT parameters \cite{28}  including normalized 
anisotropy \emph{A}, polarizance \emph{b}, circular 
birefringence $\upbeta $, and anisotropy degree $t_{1}$,
Mueller matrix rotation invariant parameters \cite{29}  
including linear diattenuation $D_{L}$, linear polarizance
$P_{L}$, circular diattenuation $D_{C}$, circular 
polarizance $P_{C}$, linear birefringence related $r_{L}$ 
and $q_{L}$, and $k_{C}$ with different physical meanings 
in pure depolarization and linear retarder system, Mueller 
matrix linear birefringence identity parameters ($P_{1}$, 
$P_{2}$, $P_{3}$, and $P_{4}$) based on the Mueller matrix
of linear retardance, and Mueller matrix linear 
diattenuation identity parameters ($P_{5}$, $P_{6}$, $P_{7}$, and $P_{8}$) based on the Mueller matrix of linear diattenuation \cite{20,42}. In this study, we used PBPs composed of above parameters as the input polarization features of the classifiers for the classification of HCC cell structures, ICC cell structures, and non-cancerous structures.  
\subsubsection{Radiomics Features}
Radiomics is a comprehensive method of medical image analysis to improve the diagnostic, prognostic, and predictive accuracy. A total of 93 radiomic features \cite{39} were used in this study. The same as the polarization features, radiomics features extracted from H\&E images of the liver tissues are used as the input features of the classifiers. These radiomic features of H\&E images quantified target microstructures’ characteristics and are subdivided into the two classes: intensity and texture. Intensity-based features were obtained by estimating the first order statistics of the intensity histogram, and described the distribution of pixel intensities within ROI image through commonly used metrics, including maximum, minimum, mean, standard deviation, variance, etc. Texture-based features were derived from the gray level co-occurrence matrix which describes the second-order joint probability function of an ROI image, gray level run length matrix which quantifies gray level zones in an ROI image, neighboring gray tone difference matrix which quantifies the difference between a gray value and the average gray value of its neighbors, gray level size zone matrix which quantifies gray level zones in an ROI image, and gray level dependence matrix which quantifies gray level dependencies in an ROI image. The texture-based features described the spatial distribution of grayscale values and quantified the heterogeneity. 
\subsection{Overview}
Fig. 2 outlines the steps from the input of Mueller matrix image and H\&E image of the liver tissues to the output of the classification results of HCC cell structures, ICC cell structures, and non-cancerous structures. Firstly, we obtained the liver tissue samples’ Muller matrix images under a 4 $\times $ objective and corresponding H\&E images under a 20 $\times $ objective. The ROI that maximizes inclusion of the three target microstructures in each sample are selected by pathologists. The affine transformation method \cite{43,44} was adopted for the image registration pixel by pixel between the Mueller matrix and H\&E images. The mask labelled by the pathologist on H\&E images is mapped to the corresponding polarization images at pixel level. After registration, we calculated the polarization parameters for each target pixel and radiomics features from the H\&E image blocks of 100 $\times $ 100 size around the target pixel. The polarization features and radiomics features of each target pixel were treated as the two modalities input data of the PRFFN for classification. We also investigated the performance of three ML classifiers—polarization features-based ML classifier, radiomics features-based ML classifier, and PRFFN combining the two features—as the imaging resolution reducing.
\begin{figure*}[h]
\centerline{\includegraphics[width=\linewidth]{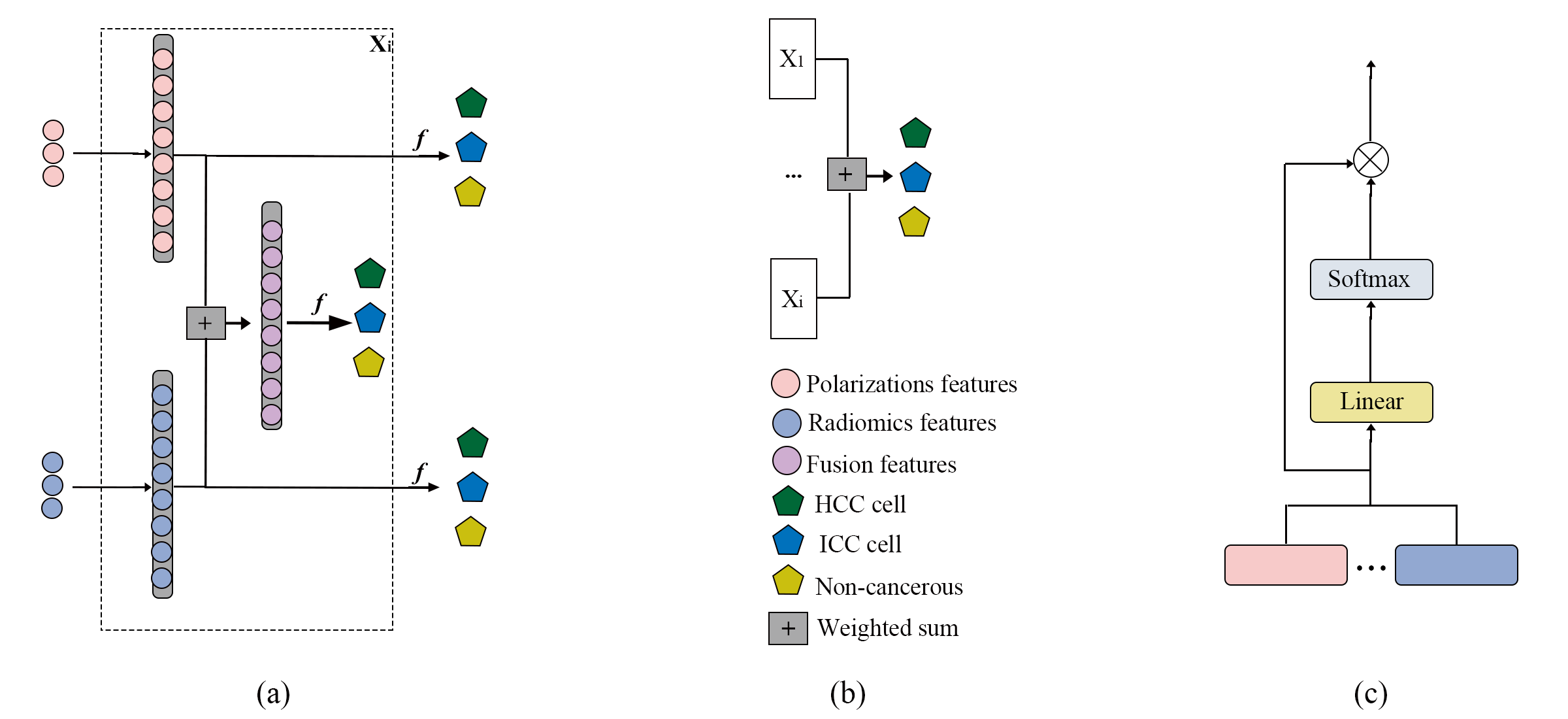}}
\caption{Architecture of the PRFFN combining polarization and radiomics features for the classification of HCC and ICC. (a) Early Fusion at the Feature Level refers to the initial stage of the model where features from different modalities (such as polarization features and image features) are combined to create a new feature representation. (b) Late Fusion at the Result Level means that prediction results from different layers and depths of features are merged to generate the final prediction. This fusion strategy considers information at multiple hierarchical levels and depths to enhance classification performance. (c) The Weighted Summation Process involves the calculation of weights, which are used to combine different modalities of features or features from different layers. This process ensures that information from different modalities and layers is appropriately balanced and fused. These components collectively constitute the structure of PRFFN, allowing it to leverage the strengths of diverse information sources for improved classification performance.}
\label{fig3}
\end{figure*}
\subsection{Data Preprocessing}

In this study, the input data of the PRFFN for classification were the target pixels with different polarization and radiomics features. In order to map the mask labelled by the pathologist on H\&E images to the corresponding polarization images pixel by pixel and extract the polarization and radiomics features of the same pixel, the affine transformation method was adopted for the image registration between the two images. The element m11 in Mueller matrix represents the intensity image of the sample. As shown in Fig. 2, the H\&E image of the sample as a moving image was transformed to match the corresponding m11 image as a fixed image for pixel level registration. In MATLAB, we conducted the registration by selecting control points common to both images and inferring the affine transformation matrix \textbf{T} that aligns the control points. After selecting control points interactively by calling the \emph{cpselect }function, the transformation matrix \textbf{T} that best aligns the moving and fixed points was produced by calling the \emph{fitgeotrans} function when the transformation type was set as “affine”. By applying the transformation matrix \textbf{T} and the H\&E images with pathologists’ masks to the m11 images using the \emph{imwarp} function, we transformed the mask to select target pixels in PBPs images as polarization features and calculated radiomics features of the H\&E images after registration, as input data of the fusion.
\subsection{Network Algorithm Architecture}

\begin{figure*}[h]
\centerline{\includegraphics[width=\linewidth]{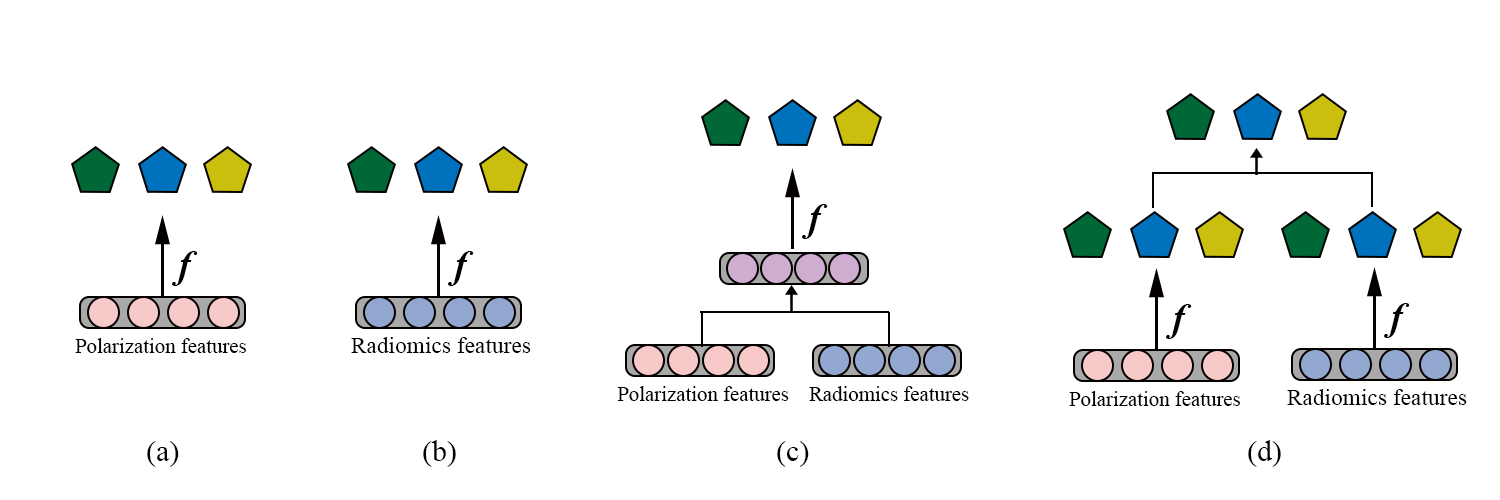}}
\caption{The other fusion methods used for comparison. (a) Using Only Polarization Features as Input: This method involves using only polarization features for classification without combining them with image features. (b) Using Only Image Features as Input: In this approach, classification is performed using only the image features from H\&E images without incorporating polarization features. (c) Early Fusion of Polarization and Image Features: This fusion strategy combines polarization features and image features at an early stage, creating a new feature representation that is used for classification. (d) Late Fusion of Results from Polarization and Image Features: In this method, the results obtained separately from the classification using polarization features and image features are combined at a later stage to produce the final prediction. These different fusion methods are employed for comparison to evaluate their impact on the classification performance and to assess the effectiveness of the proposed PRFFN in comparison to single-modality and other fusion strategies.}
\label{fig4}
\end{figure*}
As depicted in Fig. 3, the proposed fusion network comprises two levels of fusion: early feature-level fusion and late classification-level fusion. In the early feature-level fusion, we employ a weighted summation approach to combine features from both modalities, integrating polarization features and image features by learning the appropriate weights. In the late classification-level fusion, we apply a similar operation to the fused features from different layers and depths, as in the feature-level fusion. Additionally, we use a classifier to determine which layer's predictions are more crucial in order to perform the final fusion of predictions from different layers. In both levels of fusion, we utilize a simple structure involving a linear layer followed by a softmax operation. This approach not only simplifies the structure but also provides a degree of interpretability, allowing the learned weights to indicate the relative importance of different feature categories.

"In this context, assuming the presence of two types of features, with polarization features denoted as $x_{P}$ and image features as $x_{R}$, the classification results using only polarization features and image features are represented as $y_{P} = f(x_{P})$ and $y_{R} = f(x_{R})$, respectively. The classifier f employed in this context is a multi-layer perceptron (MLP). The expression for each layer in the latent space can be defined as follows:
\begin{equation}y_{Pi} =g\left (W_{Pi}x_{Pi}+b_{Pi} \right ) , i=1,\dots ,k .\label{eq}\end{equation}
\begin{equation}y_{Ri} =g\left (W_{Ri}x_{Ri}+b_{Ri} \right ) , i=1,\dots ,k .\label{eq}\end{equation}
Where $x_{Pi}$, $y_{Pi}$, $W_{Pi}$, and $b_{Pi}$ are the input, output, weight, and bias of the i-th layer in the MLP model using only polarization features. Similarly, $x_{Ri}$, $y_{Ri}$, $W_{Ri}$, and $b_{Ri}$ represent the input, output, weight, and bias of the i-th layer in the MLP model utilizing only image features. The activation function $g(h)$ employed here is the Rectified Linear Unit (ReLU), defined as $g(h) = max(0, x)$.Subsequently, early fusion is performed on the features from each layer in the latent space:
\begin{equation}y_{i} =a_{Pi}y_{Pi}+  a_{Ri}y_{Ri} .\label{eq}\end{equation}
The weights at the feature level $\left \{a_{Pi}, a_{Ri} \right \} $ are computed by the attention layer as follow:
\begin{equation}\left \{a_{Pi}, a_{Ri} \right \} =softmax\left ( Linear\left ( \left \{ y_{Pi}, y_{Ri} \right \}  \right )  \right ) .\label{eq}\end{equation}
The classification results obtained by using the fused features as input are represented as $X_{i}=f\left ( y_{i}  \right )  $. Finally, late fusion at the classification level is performed on the classification results from each layer in the latent space:
\begin{equation}Y=\sum_{i}^{k} b_{i}X_{i}   .\label{eq}\end{equation}
The weights at the classification level $\left \{ b_{1},\dots, b_{k}  \right \} $ are computed by the attention layer as follow:
\begin{equation}\left \{ b_{1},\dots, b_{k}  \right \} =softmax\left ( Linear\left ( \left \{ X_{1},\dots, X_{k}  \right \} \right )  \right ) .\label{eq}\end{equation}

The dataset for this study comprises 53 regions of measurement from HCC cells, 53 regions of measurement from ICC cells, and 42 regions of measurement from non-cancerous areas, all annotated by pathologists. From each of these cell collections, 50,000 pixels were randomly sampled for use as input data in training multiple MLP classifiers. Each pixel is characterized by a 23-dimensional polarization feature and a 93-dimensional image feature. Subsequently, we performed mean and variance normalization on each feature. The model parameters for the PRFFN were fine-tuned through grid search-based parameter optimization via cross-validation to maximize the classification accuracy of HCC cell structures, ICC cell structures, and non-cancerous structures. We implemented all classifiers using the open-source library Scikit-learn in Python version 3.8. For the MLP model using only polarization features, the hidden layer sizes were set as (512, 256, 128), with a learning rate of 0.01. Similarly, for the MLP model using only image features, the hidden layer sizes were (512, 256, 128), with a learning rate of 0.01.

\section{Results and Discussion}
\subsection{Comparison with Other Methods}
\begin{figure*}[h]
\centerline{\includegraphics[width=0.8\textwidth]{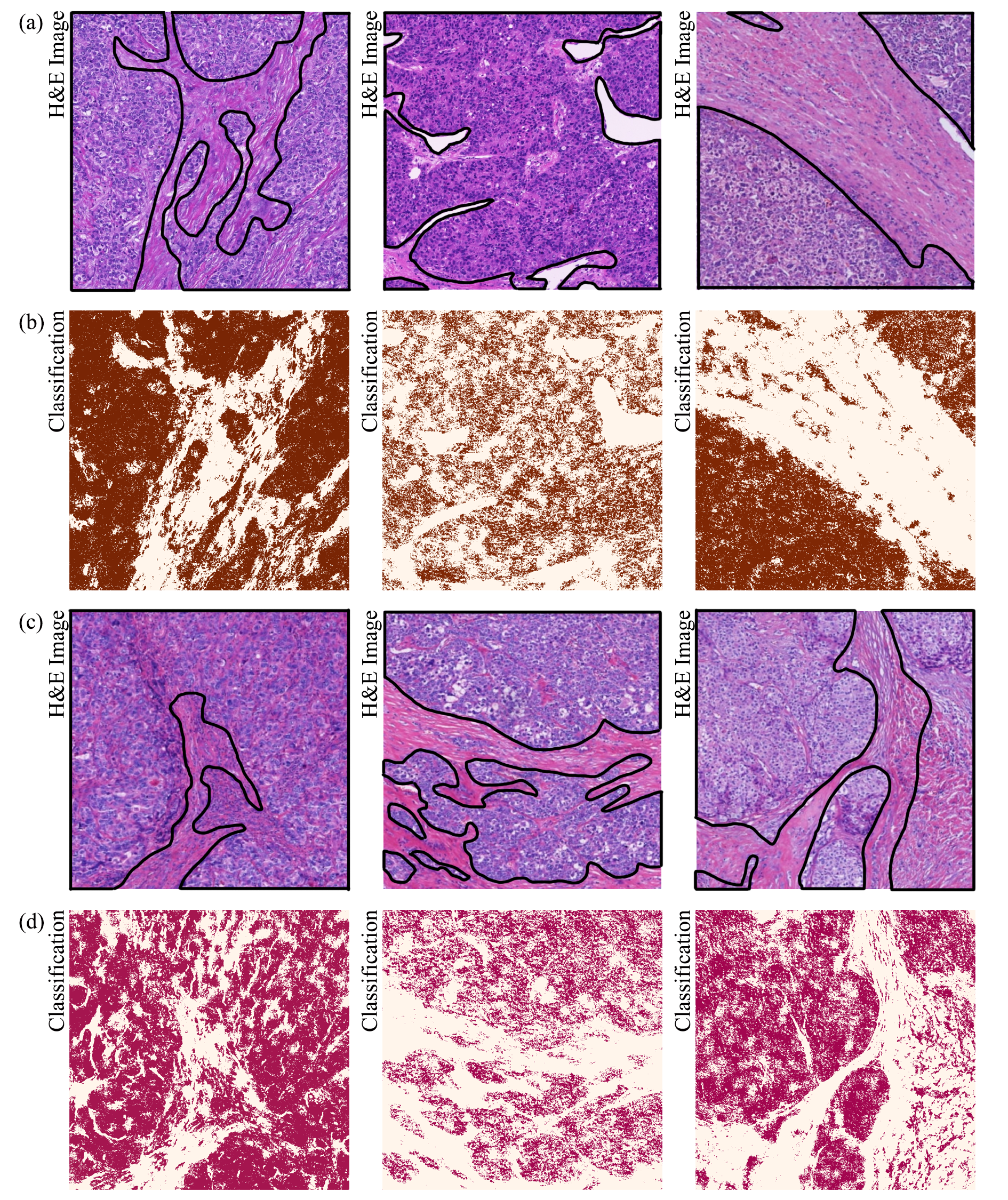}}
\caption{The characterization results of the PRFFN in the test ROIs. (a) and (c): The H\&E images of the HCC pathological slides and ICC pathological slides respectively. (b) and (d): The characterization results of the corresponding ROIs by the proposed network.}
\label{fig5}
\end{figure*}
Employing different fusion strategies for integrating features from different modalities can significantly impact the classification performance of the model. To validate the classification performance of the proposed fusion strategies and determine whether the fusion of multimodal features has the expected effects, we compared the recognition results of our designed model with single-modal classification models and other fusion strategies.

For this comparative analysis, we employed a leave-one-patient-out cross-validation approach, where the PRFFN model and other classifiers using the same input data and labels were compared through 14-fold cross-validation. As shown in Figure 3.6, in the comparison of single and dual-modal information, we compared our proposed method with the following two approaches: (1) a classifier that uses only polarization features from liver pathological tissue sections as input for HCC and ICC recognition and (2) a classifier that uses only H\&E image features from liver pathological tissue sections as input for HCC and ICC recognition. These two methods serve as single-modal feature classifiers for comparison. In the comparison of classifiers with different fusion strategies, we compared our proposed method with the following two approaches: (1) direct early fusion of polarization and image features, integrating both features to form a new feature without further feature extraction, and using the new fusion feature as input for HCC and ICC recognition and (2) late fusion of the prediction results of classifiers using only polarization features as input and classifiers using only image features as input. The results from these two modalities are combined to produce the final prediction result. These two methods serve as other fusion classifiers for comparison.

In the same manner, we considered four performance evaluation metrics: accuracy, precision, recall, and F1-score. These four metrics were employed as evaluation indicators for analyzing the classification results. During the cross-validation process, a random sample from HCC, a random sample from ICC, and a random sample from non-cancerous structures were grouped together. This process created a total of fourteen sets for classifying these three types of pathological tissue structures. In each iteration of cross-validation, one of these groups served as the test dataset, while the remaining groups were used for training. This cross-validation process was repeated 14 times. During each iteration of cross-validation, we input the true values of the test data and the features of the target microstructures into the trained model. This allowed us to predict which category each pixel in the test sample belongs to. Subsequently, we computed accuracy, precision, recall, and F1-score. Following the cross-validation process, we calculated the average values of these four metrics as the quantitative evaluation indicators. Table I summarizes the classification performance results of each network mentioned, after 14-fold cross-validation, for complex liver pathological tissue sections, including HCC cell structures, ICC cell structures, and non-cancerous structures.
\begin{table}
\caption{Performance of Different Networks on the Classification of HCC and ICC}
\label{table}
\setlength{\tabcolsep}{4.5pt}
\begin{tabular}{lllll}
\hline
                                & Accuracy & Precision & Recall & F1-score \\ \hline
Polarization feature classifier & 0.812    & 0.833     & 0.812  & 0.804    \\
Radiomics feature classifier    & 0.808    & 0.829     & 0.809  & 0.792    \\
Early fusion                    & 0.829    & 0.849     & 0.829  & 0.814    \\
Late fusion                     & 0.835    & 0.852     & 0.834  & 0.829    \\
PRFFN (our network)             & 0.877    & 0.884     & 0.877  & 0.863    \\ \hline
\end{tabular}
\label{tab1}
\end{table}

From Table I, the following conclusions can be drawn: (1) The classification performance of the dual-modal PRFFN is superior to single-modal feature classifiers. This indicates that the two types of features complement each other in the process of distinguishing HCC cell structures, ICC cell structures, and non-cancerous structures, making the structural information of pathological tissues more complete. In high-resolution H\&E-based pathological diagnosis, it is challenging for pathologists to differentiate HCC and ICC, resulting in lower accuracy for classifiers that use only H\&E image features from liver pathological tissue sections as input. On the other hand, classifiers using only polarization features from liver pathological tissue sections as input achieve higher accuracy, suggesting that the sub-wavelength microstructural polarization features, which are not visible to the human eye, play a significant role in HCC and ICC classification. (2) The proposed PRFFN outperforms classifiers that directly perform early fusion of polarization and image features, as well as classifiers that perform late fusion of prediction results from classifiers using only polarization features and classifiers using only image features. This demonstrates the superiority of the fusion approach proposed here, which combines the advantages of both early and late fusion. This fusion involves two types of fusion: feature fusion between the two modalities and fusion of prediction results from different layers and depths. It is the deep fusion involving both of these aspects that leads to higher prediction accuracy. (3) By comparing the evaluation metrics of each network, it is evident that the proposed PRFFN significantly outperforms all the other networks compared for the classification of HCC cell structures, ICC cell structures, and non-cancerous structures. It achieves an accuracy of 87.67\%, precision of 88.39\%, recall of 87.65\%, and an F1-score of 86.26\%. Using an effective fusion method that combines polarization features and H\&E image information, the classification performance of HCC and ICC is substantially improved. This indicates that PRFFN could be a powerful tool for automatic identification of the two types of cancer cells under a high-resolution microscope, eliminating the need for immunohistochemical staining and manual observation by pathologists, which could alleviate the burden on medical professionals and the diagnostic complexity to some extent.

\subsection{Quantitative Characterization Results}
ROIs selected and labelled by experienced pathologists from test samples consists of HCC cells and non-cancerous structures (as shown in Fig. 5(a)) or ICC cells and non-cancerous structures (as shown in Fig. 5(c)). The Mueller matrix images and H\&E images of the selected ROIs can be obtained by using DoFPs MMM and be calculated the polarization features and image features as input of the proposed PRFFN. Fig. 5 summarizes the output of our network in the test ROIs, and pseudo color images are presented here: the brown, red, and white pixels represent HCC cell structures, ICC cell structures, and non-cancerous structures respectively. The PRFFN determined which class each pixel in ROIs belongs to. In Fig.5 (a) and (c), the corresponding cancer cells regions were labelled by pathologists using black solid line as the ground truth of the classification. The identification results of the network in the corresponding ROIs are shown in the Fig. 5 (b) and (d), from which we can observe that: (1) In the ROIs with HCC cell structures and non-cancerous structures, the brown pixels predicted by the network can indicate the positions of HCC cells. Meanwhile, there are almost no red pixels representing ICC; (2) In the ROIs composed by ICC cell structures and non-cancerous structures, the 2D images of the output of the network have a large number of red pixels at cells positions, indicating there is few HCC cells.
\subsection{Validation of the Stability of PRFFN with Decreasing Image Resolution}
\begin{figure}[h]
\centerline{\includegraphics[width=\columnwidth]{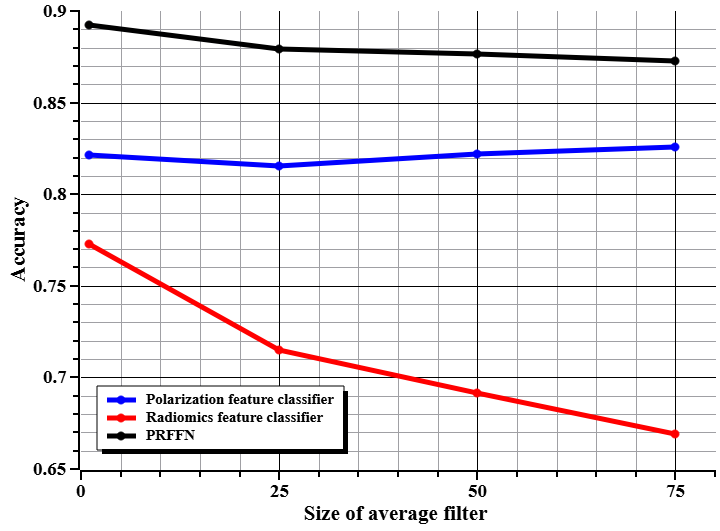}}
\caption{Accuracy on the classification of HCC and ICC at multi-resolution cases. With decreasing image resolution, the black line represents the accuracy of the PRFFN, the blue line represents the accuracy of PFMLP, and the red line represents the accuracy of RFMLP.}
\label{fig6}
\end{figure}
By sliding on the PBPs images and H\&E images with gradually increasing sizes of window of average filter, the image resolution decreases gradually. We evaluated the accuracy of classifiers that use only polarization features from liver pathological tissue sections as input for HCC and ICC recognition, classifiers that use only H\&E image features, and the PRFFN on multi-resolution cases in 12 patients. As shown in Fig. 6, for the classification of HCC cell structures, ICC cell structures, and non-cancerous structures, the accuracy of classifier that uses only H\&E image features significantly decreases from 77.3\% to 66.9\% with decreasing resolution of H\&E images. At the same time, that of classifiers that use only polarization features remains stable with decreasing resolution of PBPs images. It means that classifiers that use only polarization features take full advantage of polarization imaging whose imaging mechanism depends on each pixel’s polarization characteristics and less on imaging resolution and which could provide effective information that may not be visible to human eyes. Due to the addition of polarization information, which performs well and steadily in classifying the target microstructures at multi-resolution cases, the accuracy of the PRFFN was reduced from 89.3\% to 87.3\%, which is obviously less compared with that of classifier that uses only H\&E image features. It proved that the proposed PRFFN has a relatively stable and satisfactory performance on the classification of HCC and ICC under low image resolution cases, which paves the way for automated and rapid screening of different liver cancer cells in a low-resolution and wide-field system.

\section{Conclusion}
In this paper, we proposed a dual-modality PRFFN to fuse polarization features derived from low-resolution Mueller matrix images and radiomics features derived from high-resolution H\&E images and demonstrated the application potential of Mueller matrix microscopy in the classification of HCC and ICC. Muller matrix image is a complete description of sample polarization characteristics, which contains rich information on the microstructure and optical properties of samples. Radiomics features enable quantifying the relationship between pixels and spatial distribution of structure on H\&E images. The technique takes advantage of the sub-wavelength microstructural information reflected by polarization features at each pixel and spatial structure information decoded by radiomics features. We input polarization features and radiomics features to the PRFFN and output the classification of HCC cell structures, ICC cell structures, and non-cancerous structures at each pixel. In the designed fusion model, two levels of fusion are incorporated. Initially, there is an early fusion of features from different modalities, followed by the fusion of prediction results from different layers. The experiment results show that the classification performance of our proposed network is superior to that of a single-modality feature classifier or dual-modality fusion network with other fusion strategies. Especially, when reducing the resolution of H\&E images, the classification performance of the PRFFN remains stable and satisfactory due to the addition of polarization features. This technique provides a potential tool for computer-aided diagnosis of HCC and ICC on H\&E pathological samples, paves the way for automated and rapid screening of different liver cancer cells under a low-resolution and wide field system, and demonstrates the necessity and advantages of integrating polarization imaging methods into current image-based digital pathological diagnosis.

\bibliographystyle{IEEEtran}
\bibliography{ref.bib}
\end{document}